\newif\ifarxivpreprint
\arxivpreprinttrue      

\documentclass{article}
\usepackage{ijcai26}

\usepackage{times}
\usepackage{soul}
\usepackage{url}
\usepackage[hidelinks]{hyperref}
\usepackage[utf8]{inputenc}
\usepackage[small]{caption}
\usepackage{graphicx}
\usepackage{amsmath}
\usepackage{amssymb}
\usepackage{amsthm}
\usepackage{booktabs}
\usepackage{multirow}
\usepackage{algorithm}
\usepackage{algorithmic}
\usepackage[switch]{lineno}
\usepackage{tikz}
\usetikzlibrary{positioning, arrows.meta, calc}
\usepackage{xcolor}

\definecolor{ec_unc}{HTML}{7D7D7D}
\definecolor{ec_app}{HTML}{1D9E75}
\definecolor{ec_mag}{HTML}{1F77B4}
\definecolor{ec_rev}{HTML}{9C5BCB}

\definecolor{nfill_y2}{HTML}{BDCDF2}
\definecolor{nfill_rny}{HTML}{A8E6CF}
\definecolor{nfill_tp}{HTML}{D4A3F7}
\definecolor{nfill_be}{HTML}{FAC9C4}
\definecolor{edge_qe}{HTML}{1F77B4}
\definecolor{edge_lo}{HTML}{D62728}

\ifarxivpreprint
  \usepackage{orcidlink} 
  \usepackage{fancyhdr} 
  \usepackage{lastpage} 

  \def\PAGENUMS{1} 
  \def\PUBSTATUS{1} 

  \newcommand{\VenueStatus}{%
  Selected for Long Oral presentation at FinLLM@IJCAI 2026 (non-archival)}
\else
  \usepackage[switch]{lineno}
  \linenumbers
\fi

\title{Enhancing Regime Shift Detection Using Unstructured Data:\\
A Study on the Treasury Market}

\ifarxivpreprint
    \author{
    Mingxuan Yi\,$^1$\,\orcidlink{0009-0003-6264-9358}
    \and
    Vidal Mehra\,$^2$
    \and
    Jing Chen\,$^3$\,\orcidlink{0000-0001-7135-2116}
    \And
    John Cartlidge\,$^1$\,\orcidlink{0000-0002-3143-6355}
    \\[0.5em]
    \affiliations
    {\small
    \begin{tabular}{c}
        $^1$School of Engineering Mathematics and Technology, University of Bristol, UK \\
        $^2$Propellant Digital B.V., Amsterdam, Netherlands \\
        $^3$School of Mathematics, Cardiff University, UK
        \end{tabular}
    }\\[0.25em]
    \emails
    {\footnotesize
    mingxuan.yi@bristol.ac.uk,
    vidal.mehra@propellant.digital,
    ChenJ60@cardiff.ac.uk,
    john.cartlidge@bristol.ac.uk
    }
    }
\else
    \author{
    Mingxuan Yi$^1$
    \and
    Vidal Mehra$^2$\and
    Jing Chen$^{3}$\And
    John Cartlidge$^1$\\
    \affiliations
    $^1$School of Engineering Mathematics and Technology, University of Bristol, UK\\
    $^2$Propellant Digital B.V., Amsterdam, Netherlands\\
    $^3$School of Mathematics, Cardiff University, UK\\
    \emails
    mingxuan.yi@bristol.ac.uk,
    vidal.mehra@propellant.digital,
    ChenJ60@cardiff.ac.uk,
    john.cartlidge@bristol.ac.uk
    }
\fi

\begin{document}

\maketitle
\pagestyle{plain}
\thispagestyle{plain}

\ifnum\PAGENUMS=1
    \thispagestyle{fancy}
    \pagestyle{fancy}
    \fancyfoot[C]{\fontsize{8}{10} \selectfont Page \thepage ~of {\hypersetup{hidelinks}\pageref{LastPage}}}
    \fancyhead[L,C,R]{} 
    \setlength{\headheight}{14pt}
    \setlength{\headsep}{18pt}
    \ifnum\PUBSTATUS=1 
        \fancyhead[C]{\fontsize{8}{10} \selectfont \VenueStatus}
    \else
        \renewcommand{\headrulewidth}{0pt} 
    \fi
\fi

\begin{abstract}

Regime shifts in financial markets reorganise the joint dynamics of asset prices and macro variables, 
breaking any single-regime calibration. They are nonetheless hard to identify: the data signal is noisy
and heavily multicollinear, while the contemporaneous text that announces them is unstructured. 
Standard regime shift detection reads only the data panel and ignores this text, 
even though it typically signals the shift weeks before it materialises in observed prices. 
We address this with a text-enhanced pipeline that cross-validates the two signals. 
A large language model (LLM) proposes candidates from text, which a likelihood-ratio vector-autoregression (VAR) test 
validates on the panel. In parallel, any regime shift detector proposes data-side candidates that a second LLM call accepts via a
permissive text check. Because the acceptance stage consumes a candidate set rather than an algorithm's internals, 
the data channel accepts any data-driven detector. 
We deploy the pipeline on the US Treasury market, pairing 2010--2024 FOMC minutes with a 14-variable Treasury / macro panel, 
with every method evaluated on this same panel. The pipeline reaches $F_1$ = 0.82 and $F_2$ = 0.86 against a
verified anchor list of monetary-policy regime shifts (best with rolling PCMCI as the data channel), with same-day modal
detection latency, and is detector-agnostic: any of four interchangeable data-driven detectors clears the strongest pure data-only baseline on both scores.

\end{abstract}
\section{Introduction}
Financial time series exhibit reproducible regime shifts: discrete intervals during which the data-generating process is approximately stable, separated by transitions arising from external structural change. US monetary policy makes the examples concrete. Around the December-2015 lift-off, the Federal Reserve raised rates for the first time since the global financial crisis and exited the zero lower bound: the 2010--2015 quantitative-easing regime, in which the Fed-controlled term premium drove the yield curve, gave way to dynamics increasingly set by market-driven inflation expectations. Conversely, between March 2020 and early 2022 the policy rate returned to the zero lower bound under the COVID-19 emergency response, with large-scale asset purchases (quantitative easing) again the binding instrument: term premium, dealer balance-sheet capacity, and forward-guidance text drove yield-curve dynamics while the conventional short-rate channel was effectively shut.

\begin{figure}[t]
\centering
\scalebox{0.78}{%
\begin{tikzpicture}[
    font=\scriptsize,
    >=Latex,
    rsmnode/.style={rectangle, rounded corners=3pt, draw=black!75, line width=0.5pt, inner sep=2pt, minimum width=1.7cm, minimum height=0.78cm, font=\tiny, align=center},
    n_y2/.style ={rsmnode, fill=white},
    n_rny/.style={rsmnode, fill=white},
    n_tp/.style ={rsmnode, fill=white},
    n_be/.style ={rsmnode, fill=white},
    root/.style ={draw=black!95, line width=1.4pt},
    iso/.style  ={draw=black!30, dashed, text=black!55},
    edge_qe/.style={->, color=edge_qe, line width=1.4pt},
    edge_lo/.style={->, color=edge_lo, line width=1.4pt}
]
\begin{scope}
\node[n_y2]        (y2_1)  at (0,    0)    {2y Yield\\(policy proxy)};
\node[n_rny]       (rny_1) at (2.6,  1.3)  {10y Rate\\Expectations};
\node[n_tp,  root] (tp_1)  at (2.6, -1.3)  {Term Premium};
\node[n_be]        (be_1)  at (5.2,  0)    {Breakeven\\10y inflation};
\draw[edge_qe] (tp_1) -- (y2_1);
\draw[edge_qe] (tp_1) -- (rny_1);
\draw[edge_qe, dashed] (tp_1) to[bend right=18] (be_1);
\node[align=center, font=\scriptsize\bfseries] at (2.6, -2.15) {Regime 1: ZLB phase (Before 2015)};
\end{scope}
\begin{scope}[yshift=-4.4cm]
\node[n_y2]        (y2_2)  at (0,    0)    {2y Yield\\(policy proxy)};
\node[n_rny]       (rny_2) at (2.6,  1.3)  {10y Rate\\Expectations};
\node[n_tp,  iso]  (tp_2)  at (2.6, -1.3)  {Term Premium};
\node[n_be,  root] (be_2)  at (5.2,  0)    {Breakeven\\10y inflation};
\draw[edge_lo] (be_2) -- (y2_2);
\draw[edge_lo] (be_2) -- (rny_2);
\node[align=center, font=\scriptsize\bfseries] at (2.6, -2.15) {Regime 2: First Hike (After 2015)};
\end{scope}
\end{tikzpicture}%
}
\caption{Lagged bond-market dynamics differ qualitatively around the December-2015 Fed lift-off (the Federal Reserve's first post-2008-crisis rate hike). The lagged-DAG root recovered by rolling PCMCI swaps from \textcolor{edge_qe}{\textbf{Term Premium}} (Regime 1, the QE-era control variable) to \textcolor{edge_lo}{\textbf{Breakeven 10y inflation}} (Regime 2, the Taylor-rule input), with no robust lagged edge shared across the two regimes. Solid arrows are robust direct lagged edges, and the dashed arrow in Regime 1 is an indirect path via other factors.}
\label{fig:scm}
\end{figure}
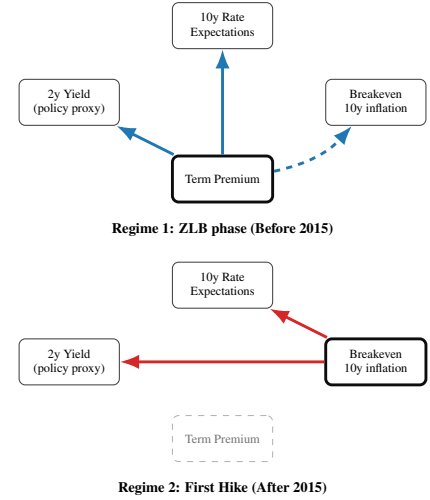

The phenomenon is not confined to these episodes: the stock--bond correlation flipped sign as inflation re-emerged in 2022~\cite{campbell-pflueger-viceira:stock-bond}, 
and the canonical flight-to-quality channel from equities to long-dated Treasuries broke down within days in March 2020~\cite{he-nagel-song:treasury-march2020}. 
Each is, in time-series terms, a structural break in the data-generating process. Locating these breaks accurately is foundational to understanding financial markets: 
the interactions among yields, term premium, and macro factors are stable, and therefore estimable, only within a single regime, so a misplaced boundary blends two distinct 
dependence structures into one. For example, Figure~\ref{fig:scm} shows that around the December-2015 lift-off the bond market's lagged dependence structure reorganises outright: 
the directed-acyclic-graph (DAG) root recovered by rolling PCMCI~\cite{runge-etal:pcmci} swaps from the Fed-controlled term premium of the quantitative-easing regime to market-driven inflation expectations. 
The problem we address is therefore detecting the timestamps at which such regime shifts occur accurately using all available signal.

Regime shift detection in multivariate time series is a mature subfield, spanning Bai--Perron structural-break testing~\cite{bai-perron}, PELT~\cite{killick-pelt}, Markov-switching VARs~\cite{sims-zha:mswar}, and, in finance, regime-aware causal-discovery methods~\cite{regime_detection_finance,saggioro-etal:regimepcmci} (see \cite{truong-cpd-review} for a survey). All of these methods rely entirely on the data signal, which is a structural limitation for treasury market data: the daily series are noisy and heavily multicollinear, since adjacent-maturity yields, the additive term-premium and risk-neutral components, and curve factors that are linear functions of the yields all co-move. The data-side fingerprint of a true regime shift is therefore easily masked by everyday co-movement that any $L_2$-cost or likelihood-based detector must absorb as noise.

For regime shifts driven by external policy decisions, such as central-bank monetary-policy changes, an informative source of evidence is available before, during, and after the shift: contemporaneous text. FOMC minutes,\footnote{The Federal Open Market Committee (FOMC) is the monetary policymaking committee of the U.S. Federal Reserve, responsible for setting interest rates and guiding U.S. monetary policy through decisions on inflation, employment, and financial conditions.} central-bank speeches, and policy communications explicitly state new policy stances, often weeks before the new dynamics materialise in observed yields, yet none of the standard detectors uses this signal. Following a collective-intelligence approach~\cite{ai4ci-position-2024-fullnames}, combining complementary information from multiple sources can provide a richer view of regime change than any single modality in isolation.

We propose a text-enhanced regime shift detection pipeline that combines a text channel (LLM-based proposals from FOMC minutes) 
with a data channel (Treasury/macro panel data), with bidirectional cross-modal validation: text-proposed candidates pass a 
data-side likelihood-ratio test, 
and data-side candidates from any regime shift detector pass an LLM permissive-prompt text check. 
The data channel is detector-agnostic: the acceptance stage consumes a candidate set, 
not a particular algorithm's internals, so any data-driven regime shift detection method can fill it.
We evaluate four (PELT, binary segmentation, Bai-Perron, and rolling PCMCI). The LLM is restricted to a regime-level decision where text carries clear signal advantage. 

\paragraph{Contributions.} (i) A \emph{detector-agnostic} text-enhanced regime-detection pipeline: LLM-based text proposals on FOMC minutes are validated by a likelihood-ratio test on the Treasury/macro panel, while the candidate set of any data-driven regime shift detection method is accepted through a permissive cross-modal text check, a framework we demonstrate with four interchangeable detectors. (ii) An empirical evaluation on 2010--2024 Federal Reserve communications paired with U.S. Treasury data, with every method scored on one consistent 14-variable panel. The text channel alone attains $F_1$ = 0.82 against a 26-event verified anchor list with same-day modal detection latency. The cross-validation pipeline matches it on $F_1$ (up to 0.82) and, under a recall-weighted $F_2$ score, exceeds it (0.86 against 0.79), while clearing the strongest pure data-only baseline (0.68) on both metrics. Text-side and data-side errors are complementary, which is what motivates the cross-validation design.

\section{Related Work}
\label{sec:related}

\paragraph{Regime shift detection:} Detecting structural breaks in multivariate time series has a long history. 
The Bai--Perron framework~\cite{bai-perron} tests for and dates multiple structural breaks in a regression. 
PELT~\cite{killick-pelt} detects change points by penalised cost minimisation and is now a standard baseline. Markov-switching VARs~\cite{sims-zha:mswar} 
infer regime labels jointly with the per-regime dynamics. Truong et al.~\cite{truong-cpd-review} survey the offline regime shift detection literature. 
Applied to multi-regime financial data, all these methods derive their signal from data alone and ignore contemporaneous policy text.

\paragraph{Regime-aware causal discovery:} CD-NOD~\cite{huang-etal:cdnod} treats nonstationarity as a surrogate variable and Regime-PCMCI~\cite{saggioro-etal:regimepcmci} jointly estimates discrete regime labels and per-regime causal structure via EM. Sadeghi et al.~\cite{regime_detection_finance} adapt CD-NOD to a lagged time-series setting (CD-NOTS) and apply it to financial markets. Although the primary contribution of this line is per-regime structure recovery, the regime-detection sub-problem is identical to ours and we treat the regime-detection output of these methods as direct baselines. We focus on detection in this paper, and downstream per-regime structure recovery is left to future work.

\paragraph{LLMs in causal and time-series reasoning:} Initial work used LLMs as priors over causal edges~\cite{kiciman-etal:llm-causal,long-etal:can-llm-build,vashishtha-etal:llm-guided}. Subsequent work moves toward LLM-error-resistant interfaces: COAT~\cite{liu-etal:coat} pairs LLM-driven factor proposal with statistical structure learning, and Vashishtha et al.~\cite{vashishtha-etal:causal-order} show that LLM ordering queries are more reliable than pairwise edge-direction queries. Ze\v{c}evi\'{c} et al.~\cite{zecevic-etal:causal-parrots} document that LLMs frequently pattern-match on memorised content rather than reason causally. We design accordingly: the LLM decides only the regime-level question, where text carries direct signal advantage, and never decides edges.
\paragraph{LLMs in finance:} Domain-adapted financial LLMs such as BloombergGPT~\cite{wu2023bloomberggpt} and the open-source FinGPT~\cite{yang2023fingpt} (surveyed by Li et al.~\cite{li2023llmfinance}) show that large models capture finance-specific language. A parallel line uses general-purpose LLMs to extract predictive signal directly from financial text: Lopez-Lira and Tang~\cite{lopezlira2023chatgpt} report that ChatGPT sentiment on news headlines forecasts subsequent returns, and recent IMF work applies LLMs at scale to central-bank communication~\cite{imf2025cbtext}. These approaches treat the LLM as a forecaster or sentiment extractor. We instead restrict it to a regime-boundary judgement and cross-validate that output against a statistical structural-break test, so it is checked rather than trusted.

\section{Enhancing Regime Detection From Text Data}
\label{sec:method}

A regime change in monetary policy can manifest in two complementary ways: (i) as a \textit{narrative shift} in central-bank communication that announces or pre-commits to a new policy stance, and (ii) as a \textit{structural break} in the joint dynamics of yields, spreads, and macro variables. Statistical detection on time-series panels alone is sensitive to (ii) but blind to (i) when the narrative leads the data, while a purely text-based detector misses regime changes that originate outside the central-bank corpus (e.g., between-meeting speeches, financial-stress events). Our pipeline reconciles both signals through a bidirectional cross-validation design.

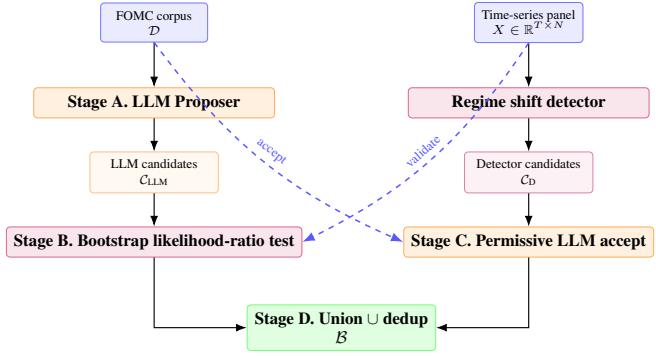
\begin{figure}[t]
\centering
\resizebox{\columnwidth}{!}{%
\begin{tikzpicture}[
    font=\small,
    node distance=4mm,
    box/.style={rectangle, draw, rounded corners=2pt, align=center, inner sep=4pt},
    input/.style={box, fill=blue!8, draw=blue!55, font=\scriptsize, minimum width=2.0cm, minimum height=0.7cm},
    llm/.style={box, fill=orange!12, draw=orange!65, minimum width=4.5cm, align=center},
    stat/.style={box, fill=purple!10, draw=purple!60, minimum width=4.5cm, align=center},
    output/.style={box, fill=green!12, draw=green!55, minimum width=3.5cm, font=\bfseries\small},
    flow/.style={->, >=Latex, semithick},
    cv/.style={->, >=Latex, dashed, blue!65, thick}
]
\node[input] (corpus) at (-3.5, 4.5) {FOMC corpus\\$\mathcal{D}$};
\node[input] (panel)  at ( 3.5, 4.5) {Time-series panel\\$X \in \mathbb{R}^{T\times N}$};
\node[llm]  (sA) at (-3.5, 3.0) {\textbf{Stage A.\ LLM Proposer}};
\node[stat] (pcmci) at ( 3.5, 3.0) {\textbf{Regime shift detector}};
\node[input, fill=orange!5, draw=orange!50, minimum width=2.4cm] (cA) at (-3.5, 1.7) {LLM candidates\\$\mathcal{C}_\text{LLM}$};
\node[input, fill=purple!5, draw=purple!50, minimum width=2.4cm] (cP) at ( 3.5, 1.7) {Detector candidates\\$\mathcal{C}_\text{D}$};
\node[stat] (sB) at (-3.5, 0.4) {\textbf{Stage B.\ Bootstrap likelihood-ratio test}};
\node[llm]  (sC) at ( 3.5, 0.4) {\textbf{Stage C.\ Permissive LLM accept}};
\node[output] (sD) at (0, -1.2) {Stage D.\ Union $\cup$ dedup\\$\mathcal{B}$};
\draw[flow] (corpus) -- (sA);
\draw[flow] (panel)  -- (pcmci);
\draw[flow] (sA) -- (cA);
\draw[flow] (pcmci) -- (cP);
\draw[flow] (cA) -- (sB);
\draw[flow] (cP) -- (sC);
\draw[cv] (panel.south) to[bend left=15] node[above, font=\scriptsize, sloped] {validate} (sB.east);
\draw[cv] (corpus.south) to[bend right=15] node[above, font=\scriptsize, sloped] {accept} (sC.west);
\draw[flow] (sB) |- (sD.west);
\draw[flow] (sC) |- (sD.east);
\end{tikzpicture}%
}
\caption{Pipeline architecture. An \textbf{LLM proposer} (Stage A) on FOMC text and \textbf{any data-driven regime shift detector} on the panel produce candidates cross-validated by the other modality: LLM candidates pass a likelihood-ratio test (Stage B), and detector candidates are accepted by a permissive LLM text check (Stage C, which consumes only the candidate set, so any detector substitutes). Stage D unions and deduplicates into $\mathcal{B}$. Dashed arrows mark cross-modal validation.}
\label{fig:pipeline}
\end{figure}

Concretely, the pipeline consumes (i) a dated corpus of FOMC minutes $\mathcal{D} = \{D_1, \ldots, D_M\}$ ordered by release date, and (ii) a multivariate financial time-series panel $X \in \mathbb{R}^{T \times N}$ over the same period. It outputs a set of validated regime-boundary dates $\mathcal{B}$. Figure~\ref{fig:pipeline} illustrates the four stages.

\paragraph{Stage A: LLM proposer.}
The strict prompt (Algorithm~\ref{alg:stage-a}, full text in Figure~\ref{fig:stage-a-prompt}) anchors three labels to explicit definitions. A regime is a multi-meeting stance with stable reaction function, dominant tool, and trajectory, while a \texttt{major\_pivot} is a transition between regimes (taper start, first hike, zero-lower-bound exit, framework adoption, ``transitory''$\to$``persistent'' reframing). A calibration prior of 0--1 pivots per year (FOMC base rate) discourages over-detection, and the LLM is asked to default to \texttt{incremental} under ambiguity.
\begin{algorithm}[t]
\caption{Stage A: LLM proposer}
\label{alg:stage-a}
\small
\begin{algorithmic}[1]
\REQUIRE FOMC corpus $\mathcal{D} = \{D_1,\ldots,D_M\}$; strict-prompt LLM \textsc{LLM-strict}; confidence threshold $\theta_A{=}0.6$
\ENSURE Text-driven candidate set $\mathcal{C}_\text{LLM}$
\STATE $\mathcal{C}_\text{LLM} \gets \emptyset$
\FOR{$i = 2,\ldots,M$}
  \STATE $(\mathit{verdict},\mathit{conf}) \gets \textsc{LLM-strict}(D_{i-1}, D_i)$ \quad \{$\mathit{verdict} \in$ \{\texttt{major\_pivot}, \texttt{incremental}, \texttt{no\_change}\}\}
  \IF{$\mathit{verdict} = \texttt{major\_pivot}$ \textbf{and} $\mathit{conf} \geq \theta_A$}
    \STATE $\mathcal{C}_\text{LLM} \gets \mathcal{C}_\text{LLM} \cup \{\mathrm{date}(D_i)\}$
  \ENDIF
\ENDFOR
\STATE \textbf{return} $\mathcal{C}_\text{LLM}$
\end{algorithmic}
\end{algorithm}
\begin{figure}[t]
\centering
\colorbox{black!6}{\begin{minipage}{0.93\columnwidth}\vspace{2pt}
\small
\textbf{System.} Expert macro-financial economist. Identify substantive shifts in monetary-policy stance.

\textbf{User.} Given two consecutive FOMC documents \texttt{\{prev\_text\}}, \texttt{\{curr\_text\}}, classify the transition as one of:
\begin{itemize}\setlength{\itemsep}{0pt}
\item \texttt{major\_pivot}: change of reaction function, dominant tool (rate / quantitative easing / quantitative tightening / forward guidance), or trajectory.
\item \texttt{incremental}: continuation along an established trajectory.
\item \texttt{no\_change}: operationally the same regime.
\end{itemize}
Prior: 0--1 \texttt{major\_pivot}s per year. Default to \texttt{incremental} when in doubt.

Output JSON: \texttt{\{regime\_change\_type, confidence (0-1), reasoning (1-3 sentences), key\_quote\}}.
\vspace{2pt}\end{minipage}}
\caption{Stage~A strict prompt (condensed).}
\label{fig:stage-a-prompt}
\end{figure}

\paragraph{Stage B: Bootstrap likelihood-ratio validation.}
Each LLM candidate $\tau$ is validated on the full 14-variable panel $X \in \mathbb{R}^{T \times N}$ ($N{=}14$) by a Chow-style structural-break test~\cite{chow1960} on a lag-1 Gaussian VAR,
\begin{equation}
X_t = c + \Phi X_{t-1} + \varepsilon_t, \qquad \varepsilon_t \sim \mathcal{N}(0, \Sigma).
\label{eq:var1}
\end{equation}
The test contrasts $H_0$: a single VAR(1) $(c,\Phi,\Sigma)$ governs the whole window $[\tau{-}W,\tau{+}W]$ (no regime shift at $\tau$), against $H_1$: the pre- and post-$\tau$ segments follow different VAR(1) parameters (a structural break at $\tau$). Algorithm~\ref{alg:stage-b} fits \eqref{eq:var1} on the pre-, post-, and full $W$-day windows around $\tau$ and forms the split likelihood-ratio statistic
\begin{equation}
\mathrm{LR}(\tau) = -2\bigl[\ell_\text{full} - (\ell_\text{pre}+\ell_\text{post})\bigr],
\label{eq:lr}
\end{equation}
where $\ell_\text{pre},\ell_\text{post},\ell_\text{full}$ are the fitted Gaussian log-likelihoods.

The asymptotic $\chi^2$ calibration of this statistic is invalid here. The unrestricted model releases $\mathrm{df} = N + N^2 + N(N{+}1)/2 = 315$ intercept, autoregression, and residual-covariance parameters across the boundary, yet each 90-day half-window holds only $\approx\!63$ daily observations, so the per-segment VAR is over-parameterised, overfits segment noise, and inflates $\mathrm{LR}$. We instead calibrate the null by a residual bootstrap, which additionally relaxes the Gaussian-innovation assumption of \eqref{eq:var1}: fit the pooled single-regime VAR(1) on $X_{[\tau-W,\tau+W]}$, resample its residuals i.i.d.\ to generate $B{=}500$ datasets under $H_0$, recompute the split $\mathrm{LR}$ on each, and set
\begin{equation}
p = \frac{1 + \sum_{b=1}^{B} \mathbf{1}\!\left[\,\mathrm{LR}^*_b \geq \mathrm{LR}(\tau)\,\right]}{B+1},
\label{eq:bootp}
\end{equation}
where $\mathbf{1}[\cdot]$ is the indicator function, so $p$ is the (add-one smoothed) fraction of bootstrap replicates at least as extreme as the observed statistic.
A candidate is retained iff $p \leq \alpha$. Algorithm~\ref{alg:stage-b} summarises the whole procedure.

\begin{algorithm}[t]
\caption{Stage B: Bootstrap likelihood-ratio validation of LLM candidates}
\label{alg:stage-b}
\small
\begin{algorithmic}[1]
\REQUIRE Candidate set $\mathcal{C}_\text{LLM}$ (each $\tau$ a candidate regime-shift date); panel $X$; half-window $W{=}90$d; bootstrap reps $B{=}500$; level $\alpha{=}0.05$
\ENSURE Data-validated candidate set $\mathcal{C}_\text{LLM-val}$
\STATE $\mathcal{C}_\text{LLM-val} \gets \emptyset$
\FOR{$\tau \in \mathcal{C}_\text{LLM}$}
  \STATE Compute $\mathrm{LR}(\tau)$ by~\eqref{eq:lr} (pre/post vs.\ full, window $W$)
  \STATE Fit pooled single-regime VAR(1) on $X_{[\tau{-}W,\tau{+}W]}$ \quad \{$H_0$ generator\}
  \FOR{$b = 1, \ldots, B$}
    \STATE $\mathrm{LR}^*_b \gets$ split-$\mathrm{LR}$ on a residual-bootstrap resample under $H_0$
  \ENDFOR
  \STATE $p \gets$ bootstrap $p$-value by~\eqref{eq:bootp}
  \IF{$p \leq \alpha$}
    \STATE $\mathcal{C}_\text{LLM-val} \gets \mathcal{C}_\text{LLM-val} \cup \{\tau\}$
  \ENDIF
\ENDFOR
\STATE \textbf{return} $\mathcal{C}_\text{LLM-val}$
\end{algorithmic}
\end{algorithm}

\paragraph{Stage C: Permissive cross-validation of detector candidates.}
The data channel runs a data-driven regime shift detector on the panel $X$, producing a candidate set $\mathcal{C}_\text{D}$. Stage~C consumes only this set, never the detector's internals, so any data-driven regime shift detection method can fill the slot, e.g., PELT, binary segmentation, Bai-Perron, and rolling-window PCMCI.
Each candidate $\tau \in \mathcal{C}_\text{D}$ is then accepted under a distinct, 
more permissive prompt (Figure~\ref{fig:stage-c-prompt}) that sees only the two FOMC documents straddling $\tau$, 
not the panel data. The LLM is asked whether either document contains substantive monetary-policy content
that could plausibly explain a structural break around $\tau$. Because the prompt is deliberately permissive (a positive verdict needs only plausible policy content, not a match between specific data movements and specific sentences),
it effectively tests whether the surrounding minutes carry enough policy substance to make the break credible as a regime boundary. A candidate $\tau$ is accepted iff the prompt returns \texttt{explains\_signal} true with \texttt{confidence}\,$\geq\theta_C$, giving the surviving set $\mathcal{C}_\text{D-cv}$. The single global threshold $\theta_C$ is the pipeline's only tunable hyperparameter (operating point $\theta_C{=}0.8$, full grid search in Section \ref{sec:ablation}).

\begin{figure}[t]
\centering
\colorbox{black!6}{\begin{minipage}{0.93\columnwidth}\vspace{2pt}
\small
\textbf{System.} Expert macro-financial economist analysing Federal Reserve communications. Identifies substantive shifts in monetary-policy stance and distinguishes material regime change from incremental adjustments.

\textbf{User.} A separate statistical detector flagged a structural break in financial market data on \texttt{\{cp\_date\}}. Look at the two FOMC documents \texttt{\{prev\_text\}}, \texttt{\{curr\_text\}} straddling that date and decide whether either contains substantive monetary-policy content that could plausibly explain such a market break. Be \emph{permissive}: answer yes if there is any plausible content.

Output JSON: \texttt{\{explains\_signal (true/false), confidence (0--1), explanation (1--2 sentences)\}}.
\vspace{2pt}\end{minipage}}
\caption{Stage~C permissive prompt (condensed). The system prompt is shared with Stage~A.}
\label{fig:stage-c-prompt}
\end{figure}

\paragraph{Stage D: Union and deduplication.}
Stage~D greedily clusters validated candidates within $\Delta = 14$ days, retains the earliest in each cluster, and tags each output by source set (LLM-only, detector-only, or both). The final boundary set is $\mathcal{B} = \mathrm{dedup}_{\Delta}(\mathcal{C}_\text{LLM-val} \cup \mathcal{C}_\text{D-cv})$.

\section{Experiments on Real Data}
\label{sec:experiments}

\paragraph{Setup.}
We evaluate our pipeline on Federal Reserve communications and US Treasury data over 2010--2024.\footnote{All code to reproduce the experiments and tables in this paper is available at \url{https://github.com/mingxuan-yi/regime_shift}.} The FOMC corpus contains 120 minutes documents (8 per year over 15 years) sourced from federalreserve.gov. The financial panel is a lean 14-variable bond-market panel (3{,}752 daily observations after alignment) covering eight categories: Treasury yields and one curve-shape factor; the 10-year ACM term-structure decomposition into expected-rates and term-premium components; the policy rate; three complementary inflation perspectives (headline CPI, core PCE, and 10-year breakeven); the unemployment rate; banking-funding stress (TED spread); cross-asset and bond-market implied volatility (VIX and the ICE BofAML MOVE Index); and Fed total assets, the direct quantitative-easing / quantitative-tightening framework observable. Redundant intermediate-maturity yields, derived curve-slope and curvature factors, and the FOMC target-rate duplicate are removed in preprocessing to reduce multicollinearity.

All methods, including Stage~B and the four data-only detectors (PELT, binary segmentation, Bai-Perron, and rolling PCMCI), run on the same 14-variable panel, rolling z-scored over a 252-day window. Scoring every method on this one panel removes any panel-mismatch confound.

The LLM is Anthropic's Claude Sonnet 4.6 at temperature 0.2. The likelihood-ratio-test window is $W=90$ days, and rolling-PCMCI uses a 200-day window with 20-day step.

\paragraph{Anchor list.}
Ground-truth regime boundaries are taken from a 26-event anchor list. The list was first proposed by prompting OpenAI GPT-5.5 for canonical US monetary-policy regime boundaries over 2010--2024, and then verified by the authors event-by-event against federalreserve.gov press releases and Federal Reserve Board policy timelines. Events without a primary-source confirmation were dropped, and the surviving 26 dates are those the authors agreed on. The list comprises canonical Fed actions (quantitative-easing / quantitative-tightening announcements, rate-cycle inflections, crisis interventions) together with well-documented forward-guidance and framework events (Operation Twist, the Evans rule, Flexible Average Inflation Targeting framework adoption, calendar-based guidance). Note that the proposing LLM (GPT-5.5) is distinct from the pipeline's proposer/acceptor LLM (Claude Sonnet 4.6), so the evaluation does not score a model against its own outputs.

\paragraph{Baselines.}
We compare three groups, all on the same panel. \emph{(i) Four data-only baselines}, each detector run alone: \textbf{PELT}~\cite{killick-pelt} ($L_2$-cost, BIC-scaled penalty); \textbf{BinSeg} (greedy binary segmentation, ruptures default, $L_2$ cost, BIC penalty); \textbf{Bai-Perron}~\cite{bai-perron} (dynamic-programming placement of $K{=}25$ structural breaks, grid stride $\mathrm{jump}{=}5$); and \textbf{PCMCI}, the prevailing partial-correlation approach used in finance~\cite{regime_detection_finance} (a PCMCI causal graph is estimated on each rolling window, and a candidate is emitted whenever the Jaccard distance between the edge sets of consecutive windows exceeds $0.80$). \emph{(ii) LLM only}: Stage~A proposals validated by the Stage~B residual-bootstrap likelihood-ratio test, isolating text-driven detection. \emph{(iii) Cross-Validation}, our full pipeline, whose data channel accepts any of the four detectors.

\begin{table}[t]
\centering
\scriptsize
\setlength{\tabcolsep}{5pt}
\begin{tabular}{lccccc}
\toprule
Method & R & P & $F_1$ & $F_2$ & Off \\
\midrule
LLM only          & 0.77 & 0.87 & \textbf{0.82} & 0.79 & $+3.1$ \\
\midrule
PELT              & 0.58 & 0.65 & 0.61 & 0.59 & $+10.6$ \\
BinSeg            & 0.65 & 0.71 & 0.68 & 0.66 & $+8.2$ \\
Bai-Perron        & 0.62 & 0.64 & 0.63 & 0.62 & $+14.8$ \\
PCMCI             & 0.35 & 0.69 & 0.46 & 0.38 & $-10.0$ \\
\midrule
\multicolumn{6}{l}{\textbf{$+$ LLM cross-validation} ($\theta_C{=}0.8$):}\\
\quad with PELT       & 0.92 & 0.65 & 0.76 & 0.85 & $-0.2$ \\
\quad with BinSeg     & 0.88 & 0.56 & 0.69 & 0.79 & $-2.7$ \\
\quad with Bai-Perron & 0.92 & 0.63 & 0.75 & 0.85 & $-1.0$ \\
\quad with PCMCI      & 0.88 & 0.77 & \textbf{0.82} & \textbf{0.86} & $+0.9$ \\
\bottomrule
\end{tabular}
\caption{Detection metrics vs.\ the 26-anchor list.
R/P = recall/precision; Off = mean signed offset (days; $+$ = lagging); $F_2$ is the recall-weighted F-score ($\beta{=}2$). Best $F_1$ and $F_2$ in bold. LLM only = Stage~A proposals validated by the Stage~B 14-variable
residual-bootstrap likelihood-ratio test.}
\label{tab:layer1-main}
\end{table}

\subsection{Two Channels, Complementary Strengths}
We illustrate with PCMCI, though the same complementarity holds for the other three detectors. Cross-validating the text and data channels reaches $F_1$ = 0.82 at 23/26 recall (Table~\ref{tab:layer1-main}), statistically indistinguishable from the text channel (0.82) and substantially above standalone PCMCI (0.46). This $F_1$ equivalence nonetheless understates the pipeline: because a missed regime shift is costlier than a false alarm, the recall-weighted $F_2 = \frac{5PR}{4P+R}$ is the more appropriate criterion, under which CV\,$+$\,PCMCI attains 0.86 against the text channel's 0.79. The gain arises because the two channels fail on largely disjoint events (Table~\ref{tab:per-anchor}): the text channel dates most breaks to the exact day of the FOMC release (mean offset $+3.1$ days, against $-10$ to $+15$ for the data-only detectors) but is blind to shifts that move market microstructure between meetings, such as the 2021-Q1 Summary-of-Economic-Projections hike-dot shift and the 2023-03 Silicon-Valley-Bank-era hike, which PCMCI alone recovers. Overall, embedding a data-driven detector in the cross-validation pipeline improves its regime-shift detection substantially over the detector alone, raising PCMCI from $F_1$ = 0.46 to 0.82, since the two modalities recover complementary events that neither identifies in isolation.

\begin{table*}[t]
\centering
\scriptsize
\setlength{\tabcolsep}{2.5pt}
\begin{tabular}{l l r rrrr rrrr}
\toprule
 &  &  & \multicolumn{4}{c}{Standalone detector} & \multicolumn{4}{c}{Cross-Validation ($\cup$ LLM)} \\
\cmidrule(lr){4-7}\cmidrule(lr){8-11}
Date & Event & LLM only & PELT & BinSeg & Bai-P & PCMCI & PELT & BinSeg & Bai-P & PCMCI \\
\midrule
2010-08-10 & Reinvestment of mortgage-backed-securities principal & $+0$ & --- & --- & --- & $-81$ & $+0$ & $+0$ & $+0$ & $+0$ \\
2010-11-03 & 2nd quantitative easing announced & $+0$ & $+29$ & $+36$ & $+29$ & --- & $+0$ & $+0$ & $+0$ & $+0$ \\
2011-08-09 & Calendar-based forward guidance & $+0$ & $-11$ & $-11$ & $-11$ & $+21$ & $-11$ & $-11$ & $-11$ & $+0$ \\
2011-09-21 & Operation Twist & $+0$ & --- & --- & --- & --- & $+0$ & $+0$ & $+0$ & $+0$ \\
2012-09-13 & 3rd quantitative easing announced & $+0$ & --- & --- & --- & $-30$ & $+0$ & $+0$ & $+0$ & $+0$ \\
2012-12-12 & Threshold guidance (Evans rule) & $+0$ & $+19$ & $+19$ & $+19$ & --- & $+0$ & $+0$ & $+0$ & $+0$ \\
2013-05-22 & Bernanke ``taper tantrum'' speech & --- & $-5$ & $-5$ & $-5$ & --- & $-5$ & $-5$ & $-5$ & --- \\
2013-12-18 & Taper officially begins & $+0$ & $+79$ & $+79$ & $+79$ & --- & $+0$ & $+0$ & $+0$ & $+0$ \\
2014-10-29 & 3rd quantitative easing ends & $+0$ & $-34$ & $-70$ & $-34$ & --- & $+0$ & $+0$ & $+0$ & $+0$ \\
2014-12-17 & ``Patient'' language introduced & --- & --- & --- & --- & --- & $-83$ & --- & $-83$ & --- \\
2015-12-16 & First post-global-financial-crisis hike & $+0$ & $+14$ & $-40$ & $+14$ & --- & $+0$ & $+0$ & $+0$ & $+0$ \\
2017-06-14 & Balance-sheet normalization principles & $+0$ & $-85$ & $-85$ & $-85$ & --- & $+0$ & $+0$ & $+0$ & $+0$ \\
2017-09-20 & Balance-sheet runoff announcement & $+0$ & --- & --- & --- & --- & $+0$ & $+0$ & $+0$ & $+0$ \\
2018-12-19 & Last hike / Powell hawkish-then-pivot & $+42$ & $-12$ & $-28$ & $-28$ & $-8$ & $-12$ & $-28$ & $-28$ & $-8$ \\
2019-01-30 & Powell pivot to ``patient'' & --- & --- & $+36$ & $+22$ & --- & $+0$ & $+0$ & $+0$ & $+0$ \\
2019-07-31 & First post-2018 cut & $+0$ & $+56$ & $+78$ & $+63$ & --- & $+0$ & $+0$ & $+0$ & $+0$ \\
2020-03-15 & COVID emergency cut + open-ended quantitative easing & $+0$ & $-16$ & $-16$ & $-16$ & $+9$ & $+0$ & $+0$ & $+0$ & $+0$ \\
2020-08-27 & Jackson Hole: Flexible Average Inflation Targeting adopted & $+20$ & --- & $-22$ & --- & --- & $+20$ & $+20$ & $+20$ & $+20$ \\
2020-12-16 & Outcome-based forward guidance & $+0$ & $+14$ & $+58$ & $+14$ & $+26$ & $+0$ & $+0$ & $+0$ & $+0$ \\
2021-03-17 & FOMC Summary of Economic Projections first shows 2023 hike dots & --- & $+90$ & $+90$ & $+90$ & $-7$ & $+90$ & $-33$ & $+90$ & $-7$ \\
2021-11-03 & Taper announced; ``transitory'' fades & $+0$ & --- & --- & --- & --- & $+0$ & $+0$ & $+0$ & $+0$ \\
2021-12-15 & ``Transitory'' dropped + accelerated taper & $+0$ & $+86$ & $+86$ & $+86$ & --- & $+0$ & $+0$ & $+0$ & $+0$ \\
2022-03-16 & First hike of 2022 cycle & $+0$ & --- & --- & --- & --- & $-5$ & $-5$ & $-5$ & $+0$ \\
2022-05-04 & 50-basis-point hike + balance-sheet runoff start & --- & --- & --- & --- & --- & --- & --- & --- & --- \\
2023-03-22 & Silicon-Valley-Bank-era hike & --- & --- & --- & --- & $+15$ & --- & --- & --- & $+15$ \\
2024-09-18 & First cut of 2024 cycle & $+0$ & $-65$ & $-65$ & $-1$ & $-35$ & $+0$ & $+0$ & $-1$ & $+0$ \\
\midrule
\multicolumn{2}{l}{\textbf{Hits / 26}} & \textbf{20} & \textbf{15} & \textbf{17} & \textbf{16} & \textbf{9} & \textbf{24} & \textbf{23} & \textbf{24} & \textbf{23} \\
\bottomrule
\end{tabular}
\caption{Per-anchor signed offset (days; $+$ = detection lags anchor) on the 26-anchor list, $\pm 90$-day tolerance, anchor-priority greedy assignment 
(each detection serves one anchor); ``---'' = no detection within tolerance. Bottom row: anchors hit out of 26.}
\label{tab:per-anchor}
\end{table*}

\subsection{Why Data Alone Is Not Enough}
The four data-only detectors reach $F_1$ between 0.46 (PCMCI) and 0.68 (BinSeg): they locate some structural breaks but at low precision, because adjacent-maturity yields and the additive term-premium / risk-neutral components co-move so strongly that a genuine policy break is hard to separate from everyday co-movement. The text channel alone clears the strongest of them (0.82 vs.\ 0.68, $+0.14$), and every cross-validation instantiation also exceeds it (0.69--0.82), by adding contemporaneous text candidates at a naturally calibrated frequency of roughly 1.5 per year that data-only detectors structurally cannot produce. The same holds on the recall-weighted $F_2$, where the pipeline (0.79--0.86) again clears every data-only detector (0.38--0.66). Relative to each standalone baseline, this gain is almost entirely a recall effect: wrapping a detector in the pipeline lifts its recall by $+0.23$ to $+0.53$ (PCMCI $0.35{\to}0.88$, PELT $0.58{\to}0.92$) while precision moves little (BinSeg the only notable drop), because the text channel recovers the policy-announced anchors the data detectors structurally miss.

\subsection{Interpretable False Positives}
\label{sec:fp}
Cross-Validation\,$+$\,PCMCI produces 30 detections at the reported setting, of which 23 match an anchor (precision 0.77) and 7 do not. Classified post-hoc against the same primary sources used to build the anchor list, all seven are interpretable early signals or intra-regime escalations rather than spurious noise (Table~\ref{tab:fp}). The reported precision is therefore a conservative lower bound: several ``false'' alarms are early detections of regime changes that the anchor list dates to a later announcement, and each marks a comparatively weak, anticipatory move whose exclusion from the ground truth is itself defensible. This pattern is detector-independent. The three text-side false positives are shared across all four instantiations, since the bootstrap-validated text candidates are common, and only the data-side ones differ. PCMCI contributes just four, all interpretable, whereas the data-only detectors each add 10--15 that cluster on recurring volatility and turn-of-year dates with no FOMC correlate, reflecting the everyday co-movement that depresses their precision in Table~\ref{tab:layer1-main} rather than interpretable signal.

\begin{table}[t]
\centering
\scriptsize
\setlength{\tabcolsep}{3pt}
\begin{tabular}{llll}
\toprule
Date & Source & Class & Notes \\
\midrule
2011-08-30 & PCMCI & sub-regime   & Aug-2011 easing-tools debate, pre-Operation-Twist \\
2012-08-14 & PCMCI & sub-regime   & pre-QE3 run-up \\
2019-06-19 & LLM   & early signal & dovish June FOMC, pre-July-2019 cut \\
2021-02-09 & PCMCI & sub-regime   & first 10y reflation spike \\
2021-09-22 & LLM   & early signal & pre-taper signalling, bridges to 2021-11 anchor \\
2022-01-26 & LLM   & early signal & Jan-2022 hawkish pivot, pre-March-2022 hike \\
2024-08-14 & PCMCI & sub-regime   & carry-trade unwind, pre-Sept-2024 cut \\
\bottomrule
\end{tabular}
\caption{Post-hoc classification of the 7 Cross-Validation\,$+$\,PCMCI detections that match no anchor. ``early signal'': the same shift detected at a pre-announcement FOMC. ``sub-regime'': an intra-regime escalation. None is spurious noise.}
\label{tab:fp}
\end{table}

\section{Ablation Studies}
\label{sec:ablation}

\paragraph{Detector-agnosticism and the Stage-C threshold $\theta_C$.}
Stage~C consumes only a candidate set, so the data-side detector is a free choice among data-driven regime shift detection methods. Holding Stage~A and Stage~B fixed, we run the pipeline with four interchangeable detectors on the same 14-variable panel and sweep the single global Stage-C confidence threshold $\theta_C$ (Table~\ref{tab:thetac}). Two patterns are stable across the sweep. (i) Every detector instantiation is comparable to the text-only LLM ($F_1$\,$=$\,0.82) rather than clearly above or below it: on a consistent panel the cross-modal effect is modality robustness, not an outright improvement. (ii) The detector ranking is stable across the sweep, with rolling PCMCI strongest under Claude (the best detector is model-dependent, Appendix~\ref{app:crossmodel}): its Jaccard-on-causal-structure criterion is the sparsest of the four (roughly 1 candidate per year against about 1.6 for the other detectors), so its union with the text channel adds the fewest false alarms. At the operating point reported in Table~\ref{tab:layer1-main} ($\theta_C{=}0.8$) all four instantiations exceed the strongest pure data-only baseline (BinSeg, 0.68) and Cross-Validation$+$PCMCI matches LLM only. Note that $\theta_C$ is a single global hyperparameter applied identically to every detector, not tuned per detector, and the full grid is reported only for transparency.

\begin{table}[t]
\centering
\scriptsize
\setlength{\tabcolsep}{6pt}
\begin{tabular}{lccccc}
\toprule
Cross-Validation $F_1$ & \multicolumn{5}{c}{Stage-C threshold $\theta_C$} \\
\cmidrule(lr){2-6}
data channel & 0.5 & 0.6 & 0.7 & 0.8 & 0.9 \\
\midrule
PCMCI       & 0.75 & 0.75 & 0.77 & \textbf{0.82} & 0.77 \\
PELT        & 0.73 & 0.73 & 0.73 & 0.76 & 0.76 \\
Bai-Perron  & 0.72 & 0.72 & 0.72 & 0.75 & 0.76 \\
BinSeg      & 0.65 & 0.65 & 0.65 & 0.69 & 0.76 \\
\bottomrule
\end{tabular}
\caption{Cross-Validation $F_1$ vs.\ the Stage-C confidence threshold $\theta_C$, per data-channel detector (14-variable panel; LLM only $F_1$\,$=$\,0.82 for reference). $\theta_C{=}0.8$ (bold) is the single global operating point used in Table~\ref{tab:layer1-main}; the same $\theta_C$ is applied to all rows, with no per-detector tuning.}
\label{tab:thetac}
\end{table}

\paragraph{Stage~C permissive vs strict cross-validation prompt.}
The permissive prompt in Stage~C asks whether the FOMC pair straddling a detector candidate contains any substantive monetary-policy content. A strict alternative would only accept candidates that the LLM also classifies as \texttt{major\_pivot} under Stage~A's prompt. Substituting the strict prompt for the permissive one in our otherwise-identical pipeline sacrifices the two anchors that the PCMCI instantiation uniquely recovers (the 2021-Q1 Summary-of-Economic-Projections hike-dot shift and the 2023-03 Silicon-Valley-Bank-era hike), because the LLM under the strict prompt declines to call them major pivots: they fall between FOMC meetings or are couched in policy-continuation language. Losing them pulls Cross-Validation\,$+$\,PCMCI back toward the text-only operating point, erasing the data channel's contribution. The permissive prompt is therefore an essential design choice: it broadens what the LLM will accept just enough to preserve the modality complementarity that motivates cross-validation in the first place. Finally, the pipeline does not depend on the exact wording of this prompt. We reran Stage~C with six reworded prompts (see Appendix~\ref{app:prompt-robust} for full details), including a full paraphrase, a reordering of the text, a version that adds explicit content categories, and a version with the permissiveness instruction removed. CV\,$+$\,PCMCI stayed between $0.75$ and $0.82$ on $F_1$ (and $0.80$ to $0.86$ on $F_2$) in every case, close to the resampling noise floor of the operating prompt and always above the strongest data-only baseline.

\paragraph{Robustness across LLMs.}
We repeat the entire pipeline with DeepSeek-V4-Pro in place of Claude Sonnet~4.6, holding all other components fixed (see Appendix~\ref{app:crossmodel} for full details). Every principal result is preserved: the text channel alone stays above every pure data-only baseline ($F_1$ $=0.83$, $F_2$ $=0.84$), and all four cross-validation detectors match it ($F_1$ between $0.77$ and $0.80$, $F_2$ between $0.82$ and $0.91$). The results therefore derive from the pipeline design rather than from any single model.

\paragraph{Other hyperparameters.}
$\theta_C$, swept above, is the only tunable hyperparameter in the pipeline. The remaining settings are fixed by design rather than chosen on the data. The Stage-A confidence cutoff $\theta_A{=}0.6$ is the
FOMC-base-rate prior of Algorithm~\ref{alg:stage-a}. The Stage-B residual bootstrap rejects the null for every text candidate by a wide margin on the 14-variable panel, so the likelihood-ratio window is not
a critical choice, and $W{=}90$ d and the $\pm 90$ d matching tolerance are held fixed throughout.



\section{Discussion and Limitations}
\label{sec:discussion}

\paragraph{Discussion:}
The contribution is not an outright $F_1$ improvement but a \emph{robustness} result: the text channel alone attains $F_1$ = 0.82, and wrapping any of four interchangeable data-driven detectors in the cross-validation pipeline matches that level while every pure data-only baseline reaches at most 0.68. Under the recall-weighted $F_2$, which is the more appropriate criterion when a missed shift is costlier than a false alarm, the pipeline goes further and exceeds the text channel (0.86 against 0.79 for the strongest detector). Stage~B gives LLM proposals an explicit statistical grounding (a residual-bootstrap structural-break test that stays valid under an over-parameterised VAR), and Stage~C lets an arbitrary detector contribute without exposing the rest of the pipeline to its internals.

\paragraph{Limitations:}
The data channel's marginal contribution on this panel is modest: the LLM already recovers most policy-announced anchors, so cross-validation adds only a handful of intra-meeting microstructure events. This suffices to improve $F_2$ but leaves $F_1$ unchanged, and it comes at the cost of added false alarms. Regime shifts that are not announced in text, being purely market-driven, are unreachable by the text channel by construction. The 26-event anchor list is a curated ground truth, and precision is sensitive to its completeness: several nominal false positives are interpretable early signals or genuine policy actions absent from the list. The Stage-B VAR(1) is over-parameterised at $N{=}14$ (the bootstrap repairs calibration but not power) and fires on joint mean and volatility breaks. Finally, we evaluate only the US Treasury market, chosen because its data is the most complete and liquid and its policy record (FOMC minutes) the best documented. Future work will extend the pipeline to the UK, euro-area, and Asian sovereign bond markets.

\section{Conclusion}

We presented a text-enhanced regime-shift detection pipeline for bond markets, built on a simple division of labour: the LLM is restricted to the one question where contemporaneous text carries a genuine signal advantage, namely whether a regime boundary is present, while edge-level identification is delegated to statistical methods. Because the cross-modal acceptance stage consumes only a candidate set, the data channel is detector-agnostic, and PELT, binary segmentation, Bai-Perron, and rolling PCMCI all clear the strongest pure data-only baseline once wrapped in the pipeline. On 2010--2024 Federal Reserve communications the pipeline matches the text-only channel on $F_1$ and exceeds it under the recall-weighted $F_2$, and both findings hold across prompt wordings and under a second language model. Extending the approach to other central banks and asset classes, where a dated policy corpus can be paired with a market panel, is a natural direction for future work.

\section*{Acknowledgments}
This work is funded by UKRI EPSRC grant No. EP/Y028392/1: AI for Collective Intelligence (AI4CI). 

\bibliographystyle{named}
\bibliography{ijcai26}

\clearpage
\appendix

\section{Stage-C Prompt-Robustness Ablation}
\label{app:prompt-robust}
The Stage~C acceptor relies on a single hand-written prompt, which raises the question of whether the reported performance is an artefact of its exact wording. Holding Stage~A, Stage~B, and $\theta_C$ fixed, we re-ran Stage~C under six prompt variants (Table~\ref{tab:promptrobust}):
\begin{itemize}\setlength{\itemsep}{1pt}
\item \texttt{v1}: the original prompt.
\item \texttt{v1-rs}: the same prompt resampled at temperature $0.2$, which shows how much the score moves from randomness alone.
\item \texttt{cat}: a version that adds explicit content categories and asks whether the text could \emph{accompany} rather than explain the break.
\item \texttt{para}: a full paraphrase.
\item \texttt{no-perm}: the original with the phrase ``be permissive'' removed.
\item \texttt{d-1st}: a version that reorders the prompt to put the two documents before the question.
\end{itemize}
The sweep exhibits two regularities. First, resampling the same prompt already shifts $F_1$ by up to $0.03$ (\texttt{v1} against \texttt{v1-rs}), so variation below this threshold is indistinguishable from sampling noise. The reordered (\texttt{d-1st}) prompt sits within this band and the category-augmented (\texttt{cat}) prompt just outside it, while the full paraphrase (\texttt{para}) is the largest wording effect yet still leaves CV\,$+$\,PCMCI at $0.75$. Across all six wordings CV\,$+$\,PCMCI stays between $0.75$ and $0.82$ on $F_1$ (and $0.80$ to $0.86$ on $F_2$), always above the strongest data-only baseline. Second, the wording affects precision almost exclusively and leaves recall nearly unchanged. This follows directly from the design: the text candidates are fixed, so the Stage~C prompt governs only how many data-side candidates are admitted, and the additional admissions are predominantly false alarms. The sole exception is \texttt{no-perm}. Removing the permissiveness instruction tightens acceptance, reducing PCMCI's admissions from $8$ to $5$ of $13$ candidates. This improves precision but discards the two between-meeting anchors that only the data channel recovers, reproducing the trade-off reported for the strict prompt in Section~\ref{sec:ablation}. The permissiveness instruction is therefore an essential design choice rather than an incidental hyperparameter. Only BinSeg, already the weakest detector at \texttt{v1} ($0.69$), falls marginally below the strongest data-only baseline under the most permissive wordings, to as low as $0.65$, while the remaining three detectors exceed it throughout.

\begin{table}[ht]
\centering
\scriptsize
\setlength{\tabcolsep}{4pt}
\begin{tabular}{l cc cc cc cc}
\toprule
 & \multicolumn{2}{c}{PCMCI} & \multicolumn{2}{c}{PELT} & \multicolumn{2}{c}{Bai-P} & \multicolumn{2}{c}{BinSeg} \\
\cmidrule(lr){2-3}\cmidrule(lr){4-5}\cmidrule(lr){6-7}\cmidrule(lr){8-9}
Variant & $F_1$ & $F_2$ & $F_1$ & $F_2$ & $F_1$ & $F_2$ & $F_1$ & $F_2$ \\
\midrule
\texttt{v1}     & \textbf{0.82} & 0.86 & 0.76 & 0.85 & 0.75 & 0.85 & 0.69 & 0.79 \\
\texttt{v1-rs}  & 0.79 & 0.82 & 0.77 & 0.86 & 0.76 & 0.85 & 0.71 & 0.80 \\
\texttt{cat}    & 0.78 & 0.84 & 0.74 & 0.84 & 0.73 & 0.83 & 0.67 & 0.78 \\
\texttt{para}   & 0.75 & 0.83 & 0.71 & 0.82 & 0.70 & 0.82 & 0.65 & 0.77 \\
\texttt{no-perm} & 0.78 & 0.80 & 0.79 & 0.82 & 0.81 & 0.85 & 0.78 & 0.84 \\
\texttt{d-1st}  & 0.79 & 0.85 & 0.73 & 0.83 & 0.72 & 0.83 & 0.66 & 0.78 \\
\bottomrule
\end{tabular}
\caption{Cross-Validation $F_1$ and $F_2$ under six variants of the Stage-C prompt (variant codes as listed above).}
\label{tab:promptrobust}
\end{table}

\section{Cross-Model Robustness}
\label{app:crossmodel}
To assess whether the results depend on a particular language model, we repeat the entire pipeline with DeepSeek-V4-Pro in place of Claude Sonnet~4.6 in both LLM roles (the Stage~A proposer and the Stage~C acceptor), holding all remaining components fixed. DeepSeek proposes $27$ Stage-A candidates, compared with $23$ for Claude, all of which pass the Stage-B test. Table~\ref{tab:crossmodel} compares the two models, and every principal finding is preserved. The text channel alone exceeds the strongest pure data-only baseline ($0.83$ against $0.68$). All four cross-validation detectors likewise exceed it, with $F_1$ between $0.77$ and $0.80$ (and $F_2$ between $0.82$ and $0.91$), while remaining level with the text channel. The data channel again contributes almost entirely through recall, lifting each detector well above its data-only baseline by $+0.31$ to $+0.50$ (PCMCI $0.35{\to}0.85$, Bai-Perron $0.62{\to}1.00$, recovering all $26$ shifts), which is what matters when the objective is to miss no shift. DeepSeek's text-side false positives are the same interpretable early-signal events reported for Claude in Section~\ref{sec:fp}. The one substantive difference is the identity of the best detector, rolling PCMCI under Claude and Bai-Perron under DeepSeek, so the detector ranking in Section~\ref{sec:ablation} should be read as specific to Claude. Under both models, however, every detector clears the data-only baselines.

\begin{table}[t]
\centering
\scriptsize
\setlength{\tabcolsep}{5pt}
\begin{tabular}{llcccc}
\toprule
Method & LLM & R & P & $F_1$ & $F_2$ \\
\midrule
LLM only            & Claude   & 0.77 & 0.87 & 0.82 & 0.79 \\
LLM only            & DeepSeek & 0.85 & 0.81 & \textbf{0.83} & 0.84 \\
\midrule
CV\,$+$\,PELT       & Claude   & 0.92 & 0.65 & 0.76 & 0.85 \\
CV\,$+$\,PELT       & DeepSeek & 0.96 & 0.66 & 0.78 & 0.88 \\
CV\,$+$\,BinSeg     & Claude   & 0.88 & 0.56 & 0.69 & 0.79 \\
CV\,$+$\,BinSeg     & DeepSeek & 0.96 & 0.64 & 0.77 & 0.87 \\
CV\,$+$\,Bai-Perron & Claude   & 0.92 & 0.63 & 0.75 & 0.85 \\
CV\,$+$\,Bai-Perron & DeepSeek & 1.00 & 0.67 & 0.80 & \textbf{0.91} \\
CV\,$+$\,PCMCI      & Claude   & 0.88 & 0.77 & 0.82 & 0.86 \\
CV\,$+$\,PCMCI      & DeepSeek & 0.85 & 0.73 & 0.79 & 0.82 \\
\bottomrule
\end{tabular}
\caption{Comparison of Claude Sonnet~4.6 and DeepSeek-V4-Pro on the 26-anchor list.}
\label{tab:crossmodel}
\end{table}

\end{document}